\documentclass[12pt,preprint]{aastex}

\shorttitle{\hst M4 Astrometry and the Galactic Constant $V_0/R_0$}
\shortauthors{Bedin et al.}

\begin{document}

\def\hst{{\sl HST}}
\def\magspt{$\buildrel{\rm m}\over .$}


\title{\hst~ ASTROMETRY OF M4 AND THE GALACTIC CONSTANT
$V_0/R_0$\footnote{Based on observations with the NASA/ESA {\it Hubble
Space Telescope}, obtained  at the Space Telescope  Science Institute,
which is operated by AURA, Inc., under NASA contract NAS 5-26555.}}

\author{Luigi R.\ Bedin and Giampaolo Piotto}
\affil{Dipartimento di Astronomia, Universit\`a di Padova}
\affil{Vicolo dell'Osservatorio 2, I-35122 Padova, Italy}
\email{bedin@pd.astro.it, piotto@pd.astro.it}
\author{Ivan R. King}
\affil{Astronomy Department, University of Washington}
\affil{ Box 351580, Seattle, WA 98195-1580}
\email{king@astro.washington.edu}
\and
\author{Jay Anderson\footnote{Present address:\ Department of Physics \& 
Astronomy, MS 108,  Rice  University, 6100  Main Street, Houston, TX 77005.}}
\affil{Astronomy Department, University of California}
\affil{ Berkeley, CA 94720-3411}
\email{jay@astron.berkeley.edu}

\begin{abstract}
From  multi-epoch   WFPC2/\hst~  observations  we  present astrometric
measurements  of stars in  the Galactic globular cluster M4 (NGC~6121)
and in  the foreground/background.   The presence of  an extragalactic
point source allows us to determine  the absolute proper motion of the
cluster,  and, through use  of the  field  stars  in this  region only
18$^\circ$ from the Galactic center, to measure the difference between
the      Oort          constants,     $A-B$.         We          find:
$(\mu_\alpha\cos\delta,~\mu_{\delta})_{\rm J2000}$ $=$ $( -13.21   \pm
0.35, -19.28 \pm 0.35) ~ \rm mas ~ yr^{-1}$, and $A-B$ = $V_0/R_0$ $=$
27.6 $\pm$ 1.7 $\rm km~s^{-1}~kpc^{-1}$.
\end{abstract}


\section{INTRODUCTION}
\label{intro}

It is well known that the Wide Field and Planetary Camera 2 (WFPC2) on
the   {\it  Hubble   Space   Telescope}   (\hst) provides    unequaled
high-resolution  photometry, compared with  ground-based work  or with
any  other   instrument  on astronomical   satellites.  However,  only
recently has it been shown (King et  al.\ 1998, Anderson \& King 2000)
that   WFPC2  and  ACS   astrometry---with well-dithered  observations
separated  by  a few years---allows  astrometric  accuracy superior to
that obtained from ground-based plates  and/or CCDs with a much longer
time-baseline (up to a factor of $\sim$20).

In this  work we  present an astrometric   study of the  geometrically
closest Galactic  globular     cluster   M4 (NGC  6121;     see  Table
\ref{parameter} for  its  fundamental parameters),  for which a  large
number of WFPC2  observations  exist.  These data allow  studying  the
motion of the background/foreground objects too.

%
%
\begin{table}[ht!]
\begin{center}
\begin{tabular}{cc}
\tableline 
\tableline
Parameter & Value  \\ 
\tableline 
${(\alpha, \delta)_{\rm J2000}}^\ast$ & (16$^h$23$^m$35.5$^s$, $-$26$^\circ$31$'$31$''$) \\ 
${(\ell, b)_{\rm J2000}}^\ast$        & (350$^\circ$.97, 15$^\circ$.97)                  \\ 
${R_{GC}}^\ast$                             & 5.9 kpc                 		         \\
${ R_{GP}}^\ast$                            & 0.6 kpc				         \\
${(\mu_{\alpha}\cos\delta)_{\rm J2000}}^\dagger$  & $-13.21\pm0.35$ mas yr$^{-1}$     \\
${\mu_{\delta_{\rm J2000}}}^\dagger$              & $-19.28\pm0.35$ mas yr$^{-1}$     \\
     \\
${(\mu_\ell\cos{b})_{\rm J2000}}^\dagger$         & $-23.30\pm0.35$ mas yr$^{-1}$     \\
${\mu_{b_{\rm J2000}}}^\dagger$                   & $-1.81\pm0.35$ mas yr$^{-1}$      \\
${V_{\rm r}}^\ast$                                & $ 70.5\pm0.3 $ km s$^{-1}$        \\
${r_{\rm t}}^\ast  $                              &  32$'$.49  \\
${r_{\rm c}}^\ast  $                              &   0$'$.83  \\
$ c^\ast  $                                       &      1.59  \\
\tableline
\tableline 
\end{tabular}
\end{center}
\caption{Some of the fundamental parameters of M4.  In order: equatorial
and  Galactic coordinates, distance from the  Galactic center, and the
Galactic plane, proper motions in equatorial and Galactic coordinates,
radial velocity, tidal and core radii, and concentration parameter.}
\begin{list}{}{}
\item $^\ast$ from the Harris on-line catalog:
{\sf http://physun.physics.mcmaster.ca/$\sim$harris/mwgc.dat}   (Harris
1996), as updated on June 22, 1999
\item $^\dagger$ this work 
\end{list}
\label{parameter}
\end{table}

The paper is structured as follows.  In  \S\ 2 we describe the complex
multi-epoch data; in \S\ 3  we describe the astrometric procedure, and
show how the field and the cluster stars  have been separated.  In \S\
4 we address the absolute proper motion of M4.  We devote \S\ 5 to the
information we can get on the Galaxy from the present study, and \S\ 6
summarizes our results.

The   color--magnitude  diagrams,   luminosity   functions, and   mass
functions will   be  treated in a   separate paper  (Bedin et al.,  in
preparation).

\section{OBSERVATIONS}
\label{obs}

Three fields in M4 were observed by Richer and collaborators (GO-5461)
in March--April  1995 with the WFPC2  camera of the  {\it Hubble Space
Telescope}.

The  color--magnitude  diagrams (CMDs)  extracted from  this data base
have been presented by Ibata et al.\  (1999), while the extended white
dwarf sequence is discussed by Richer et al.\  (1995, 1997).  The same
group    re-observed  the   outermost  of    the  GO-5461  fields   in
January--April 2001---with      a  very  deep   survey   (123  orbits,
GO-8679)---looking  for      the    white-dwarf turn-around    in  the
color--magnitude diagram, as caused  by H$_2$ opacity (Hansen et  al.\
2002), and repeating with a larger data base (Richer et al.\ 2002) the
work presented by Bedin et al.\ 2001 (B01) on the lower main sequence.
Here we complement their work with an astrometric study enabled by the
second-epoch observations of    their innermost fields  in  April 2000
(GO-8153, PI:\  King), to  determine high-precision proper  motions in
the outermost field.

In the following, we label the 1995 fields as  CENTER, NEAR, FAR, with
obvious  meaning.    Their  locations are   shown  in  Fig.\   1.  For
completeness, the figure also shows other observations not used in the
present work.

\begin{figure}
\epsscale{1.0}
\plotone{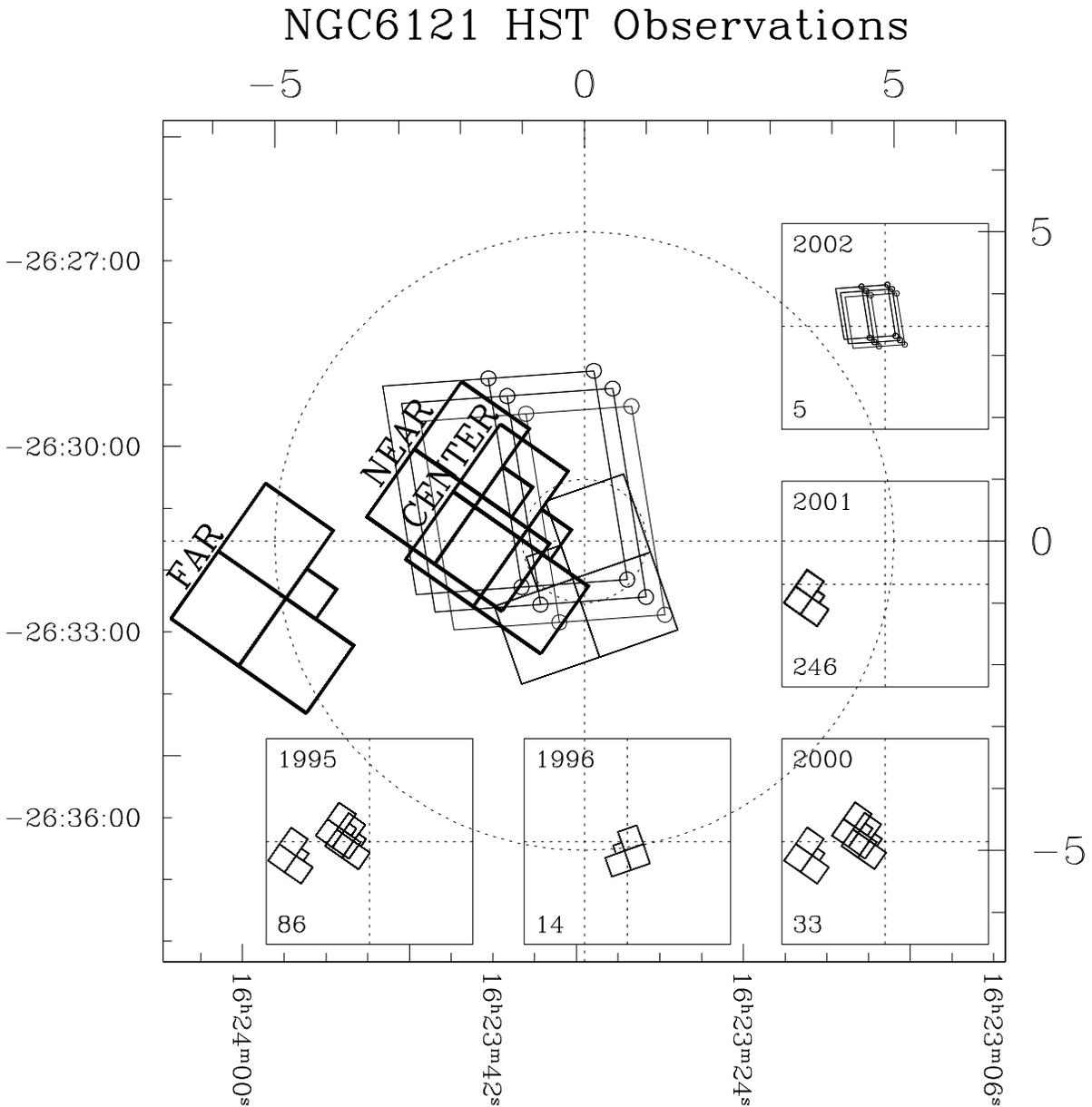}
\caption{ Finding chart of \hst~ observations of M4. Two dotted circles
are drawn at 1   and 5 arcmin  from the  center of the  cluster.   The
fields taken in 1995 have  been labeled in  this work as CENTER, NEAR,
and FAR, in  order of radial  distance from  the cluster center.   For
completeness the 1996 WFPC2 and the 2002 ACS/WFC observations are also
shown (ACS/WF1  is marked  with circles at  the corners).   The  inset
boxes   highlight   the various fields   and   epochs available.  (The
lower-left number gives the number of images.}
\end{figure}
For the CENTER and NEAR fields, the first epoch consists of 7 $\times$
1500s  $+$ 1300s in F336W, 7  $\times$ 700s $+$  600s in F814W, and 15
$\times$ 1000s in  F555W; the second epoch has  8 $\times$ 700s  $+$ 4
$\times$ 20s in  F814W.  (Short exposures  were taken in order to  get
proper motions of the brighter stars, using the first epoch in F336W.)
All sets of images are well dithered, following the recipe by Anderson
\& King (2000).  First- and second-epoch data of field FAR have already
been fully described in B01.  The third epoch for this field (GO-8679),
consists of 98 $\times$ 1300s in F606W and 148 $\times$ 1300s in F814W.
(We use the ACS Wide Field images, which consist of 5 $\times$ 360s in
F775W, only to produce the median super-sampled image in Fig.\ 5,
below.

\section{PROPER MOTIONS}
\label{pm}

We carried out  the astrometry, for  each filter and each epoch,  with
algorithms based on the effective-point-spread-function (ePSF) fitting
procedure described  by Anderson \&  King (2000).  The essence  of the
method is to  determine a finely sampled PSF   of high accuracy,  from
images at  several dither offsets.  Fitting  of this PSF to individual
well-exposed  star images  gives a  positional accuracy of  $\sim0.02$
pixel,  without any systematic errors  that depend on  the location of
the star with respect to pixel boundaries.

Unlike  B01,  where  we  used  global  transformations   between  star
positions in different epochs, here we make a local 6-parameter linear
transformation for each star, using a surrounding net of several dozen
well-measured stars  (isolated,  low residuals, cluster members,  high
signal-to-noise, not saturated), to  calculate the displacement of the
star between the two epochs.  The WFPC2  distortion has been corrected
using the new improved solution by Anderson  \& King (2003); even with
this  improved distortion correction,  however,  some distortion error
remains.  Because  its relative effect   on two points  increases with
their separation, a global   transformation has  errors that a   local
transformation avoids.  Further  details will be  given in a  paper on
our methods (Anderson \& King, in preparation).  As  a result of these
changes  in  procedure,  the  proper   motions used  here have  higher
accuracy  than  those in  B01.   Their   apparent  dispersion has  now
decreased by  about  10 \%, indicating  a  clear decrease in measuring
error.

Proper  motions have been   obtained combining the following pairs  of
data sets.  In the fields CENTER and NEAR we used F336W 1995 and F814W
2000  shallow,  for bright  stars (less   than  2 magnitudes below the
turn-off), and  F814W 1995 and F814W  2000 deep, for faint stars (more
than 2 magnitudes below the turn-off); in field FAR we used F814W 1995
and F814W 2001,  and also F555W 1995  and F606W 2001, for faint stars.
Since the purpose of this paper is purely astrometric, we will present
instrumental magnitudes only.

For convenience we will hereafter refer  to F336W, F555W, and F814W as
$U$,  $V$  and  $I$,   respectively,  although it   should  be clearly
understood that  all photometry presented here  is in the uncalibrated
instrumental pass-bands.

Figure 2 shows the vector-point diagram of relative proper motions for
all  the  independent measurements.  The axes   are  parallel to right
ascension and declination.  Since the  cluster stars have been used as
the reference list, the origin is at their centroid.

The separation   between cluster   and  field  stars  is   clear.   We
arbitrarily considered  to   be cluster members   all   the stars with
relative  proper  motion   less than  5  mas   yr$^{-1}$.    With this
membership criterion, cluster members have been  plotted in Fig.\ 2 as
small points, and the field stars as heavier points.

It is worth  noting that the difference  of motion of  the cluster and
the field  stars reflects the low tangential  motion of  M4 around the
Galactic  center  (which we shall show   below is much  lower than the
circular velocity, so that M4 is close to its apogalacticon).

\begin{figure}
\epsscale{1.0}
\plotone{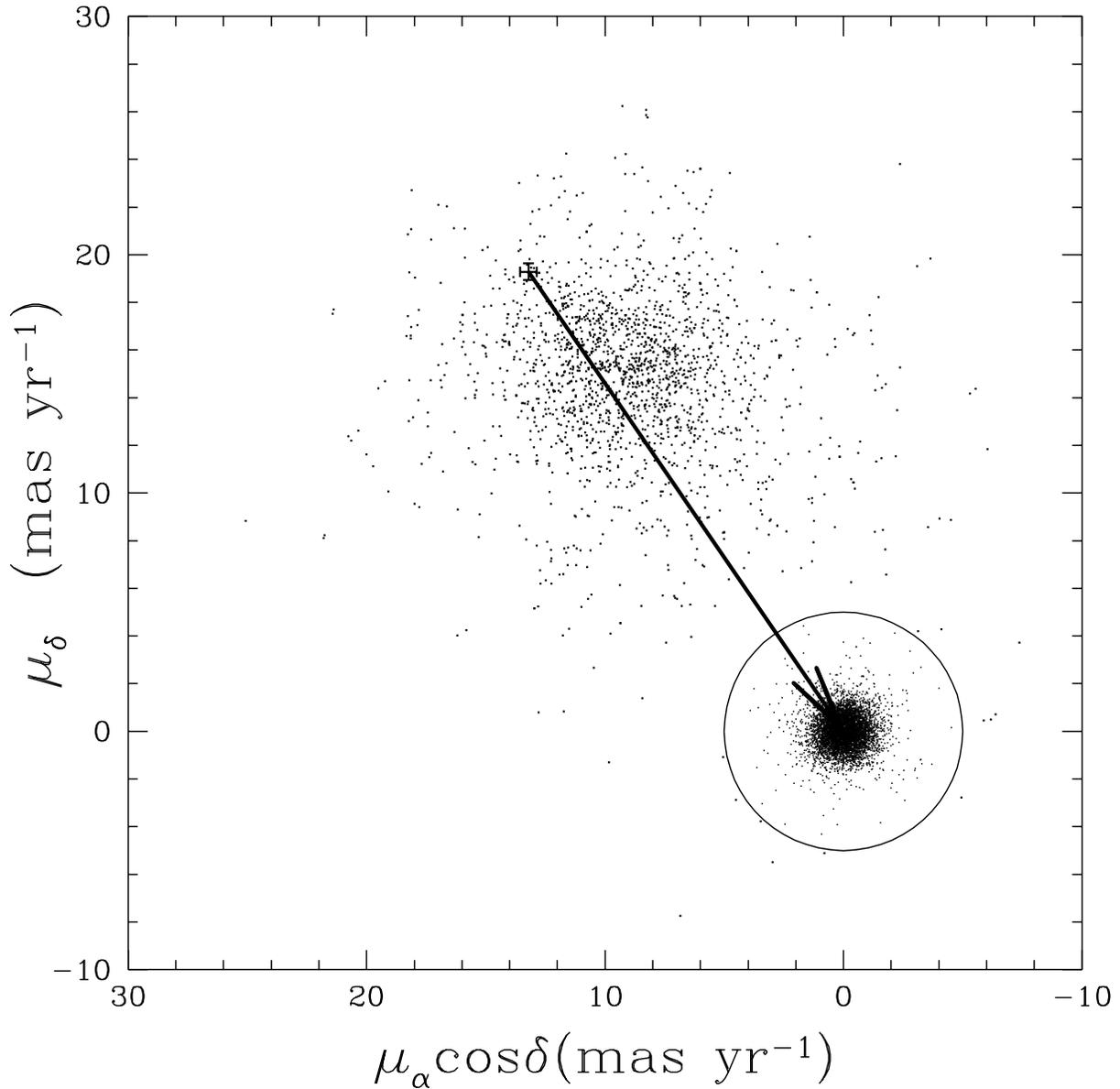}
\caption{ Vector-point diagram of all the independent 
measurements  of   the  proper    motions,  converted to    equatorial
coordinates.  Cluster members are plotted as small points, field stars
as heavier points.   Some stars appear  more than  once simply because
they have  been well measured in more  than one pair  of data sets, or
because they  fall in the  overlap region of  fields CENTER  and NEAR.
The arrow indicates the mean motion of the  cluster with respect to an
extragalactic  object whose  position   uncertainties are marked by  a
cross (see \S\ \ref{abs}).
\label{vpd} }
\end{figure}

\section{ABSOLUTE PROPER MOTION}
\label{abs}

For fields NEAR and CENTER, unlike field FAR (studied in B01), we have
F336W images, which allow us to build two-color diagrams (TCDs).

In Fig.\ 3  we show the proper-motion  separation for field NEAR.  The
top panels show the 5-year displacements in the $I$ images, and in the
bottom panels are the corresponding TCDs in ($U-V$, $V-I$) for all the
stars for  which we  have magnitudes in   all three bands.    The main
limitation in the number of stars in Fig.\  3 comes from the fact that
faint main sequence stars are missed in $U$.

\begin{figure}
\epsscale{1.0}
\plotone{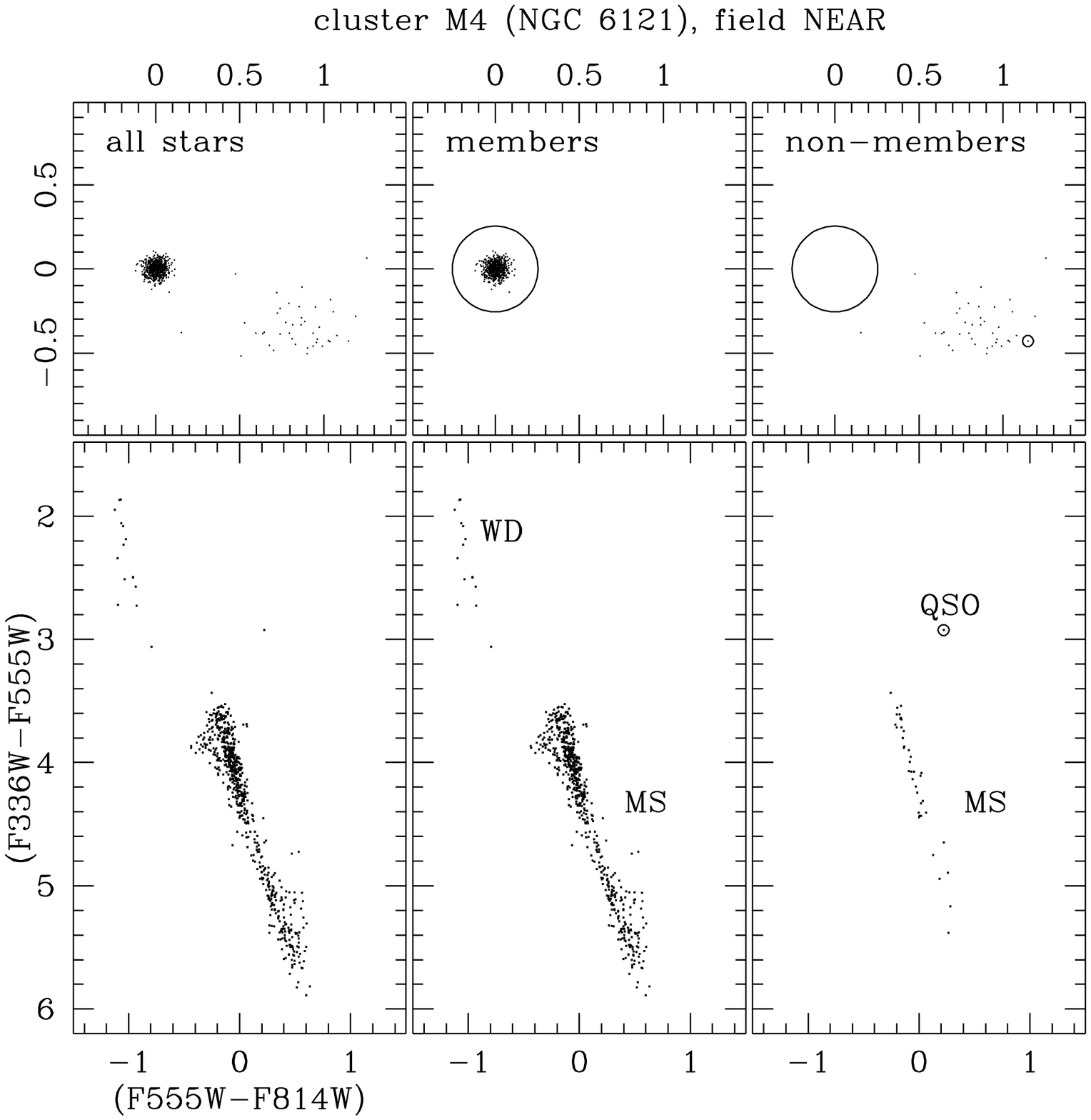}
\caption{ 
{\em Top panels:\ } 5-year displacements in the $I$ images
(instrumental unit $=$ 1 WF pixel $=$ 100 mas).
{\em Bottom panels:\ } Two-color diagrams in ($U-V$, $V-I$).
{\em Left:} All objects detected in $U$, $V$, and $I$ in field NEAR;
magnitudes are instrumental.
{\em Middle and right:\ } Separation by the proper-motion criterion
given by the circle shown.
The WD sequence and the MSs are labeled.  The extragalactic point source
is shown as an open circle.
\label{viuv} }
\end{figure}

In these  diagrams the white dwarf sequence  and main  sequence of the
cluster  are  clearly visible,  plus the  main  sequence of the field.
Note that there are no field white dwarfs in the small WFPC2 field.

An important result is the detection  of an extragalactic point-source
candidate, highlighted in the right  panels  of Fig.\  3 with an  open
circle, and labeled  ``QSO''.  This source falls  in a position in the
TCD beyond the locus defined by black bodies, incompatible with a star
but compatible with an active galactic nucleus.

To show this, in Fig.\ 4 we present a  simulated two-color diagram for
QSOs (3-pointed asterisks) and a mixture  of field stars (small dots),
kindly provided by Grazian (private  communication), superposed on the
M4 stars of  Fig.\ 4 (open squares).   The  figure shows that our  QSO
(marked by  vertical and horizontal lines) falls  in  the region where
QSOs with $z<2.8$  are most concentrated.  The  code which created the
simulated    diagram is  a modified   version   of Hyperz (Bolzonella,
Miralles, \& Pell\'o 2000) in \hst\ filters, using template stars from
Lejeune  \&  Schaerer  (2001)  and  composite  spectra  of  QSOs  from
Cristiani \& Vio (1990).

\begin{figure}
\epsscale{1.0}
\plotone{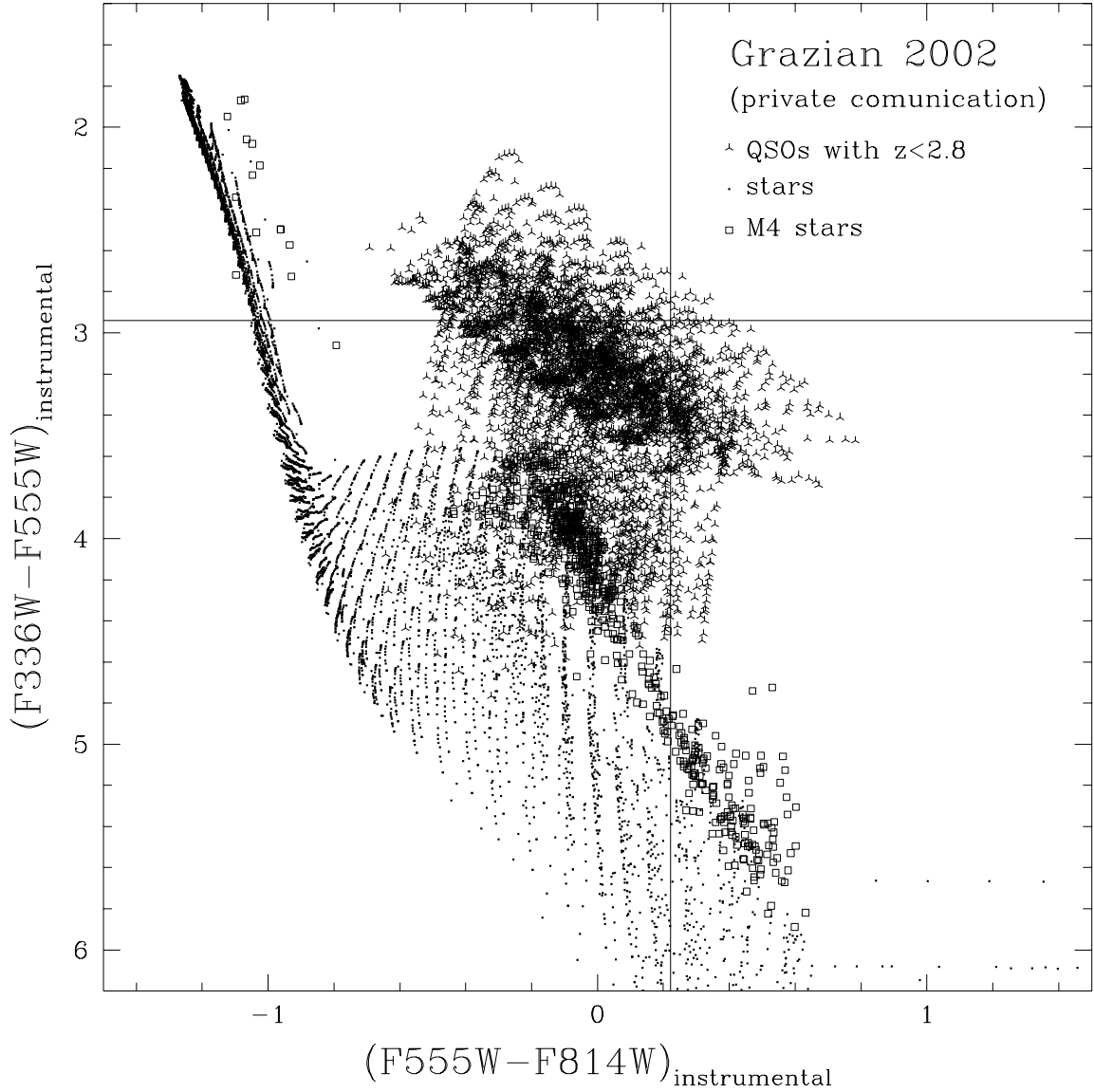}
\caption{ A simulated TCD, in \hst\ filters, of QSOs and a mixture of
field stars.  Stars are shown as dots and QSOs as 3-pointed asterisks.
The M4 stars of  Fig.\ 3 are shown  as open squares.  The intersecting
vertical and horizontal lines mark the location of our QSO.
\label{grazian}}
\end{figure}

Visual inspection of  the presumed  QSO  on an image  reveals a  faint
surrounding blur that is  not star-like.   Fig.\  5 shows a part  of a
median image of 3 F775W ACS/WFC images, super-sampled by a factor of 2
in each   direction.  (The ACS  images were  chosen  because  of their
better sampling.  These images are raw, i.e., without flat-fielding or
removal of bias and cosmic rays, but they are adequate for the present
purpose.)    In this figure, the  stars  labeled STAR1--3 are brighter
than our extragalactic source candidate (labeled QSO), but they do not
show any blur.  Though its extragalactic nature should be confirmed by
spectroscopy, the  color  and the  morphology of this  source make the
hypothesis that it is a background QSO very likely.  Its extragalactic
nature is   further supported by   the fact that   the absolute proper
motion of  M4 that we will obtain  in the following is consistent with
the  Hipparcos-based value in  the literature  (see below).  The  fact
that  this source  is very  close to point-like  allows  us to use our
effective PSF to measure its luminosity and position accurately.

\begin{figure}
\epsscale{1.0}
\plotone{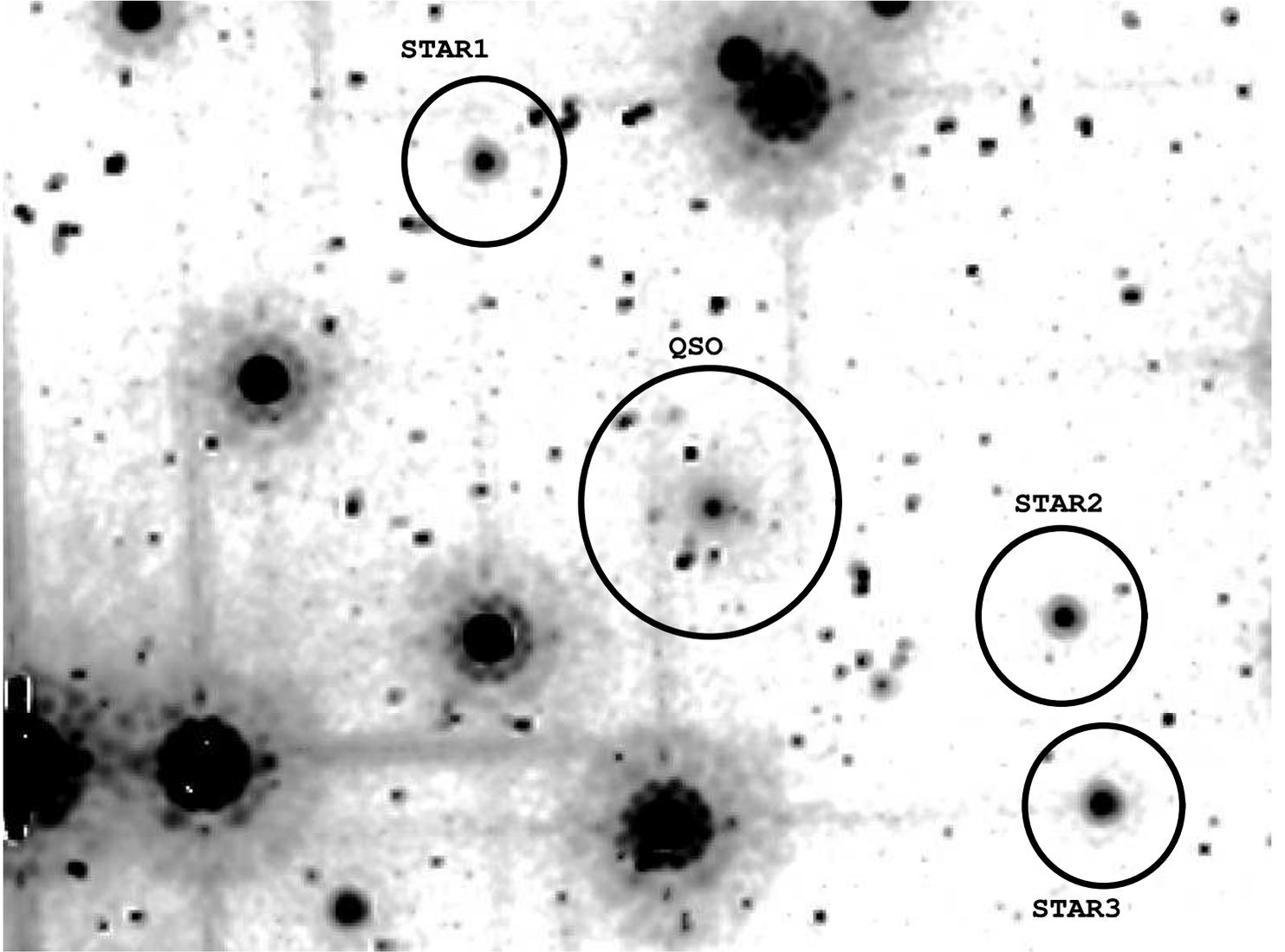}
\caption{ Super-sampled median of 3 ACS/WFC F775W images (not completely
cleaned of cosmic rays).  The extragalactic source  is labeled as QSO.
Note   how STAR1--3   (marked with  smaller    circles), although much
brighter, do not show as extended a blur.}
\end{figure}

We  can  consider the   extragalactic  point source   as  an  absolute
reference point,  from which we can infer  absolute motions of objects
in our fields.   Having derived this reference  point, we can now  use
the stars from all three of our fields, as shown in Fig.\ 2.

We find for the cluster
\begin{eqnarray}
(\mu_{\alpha}\cos\delta)_{{\rm J2000}} & = & -13.21 \pm 0.35 ~ 
\rm mas~yr^{-1}, \nonumber \\
\mu_{\delta,{\rm J2000}} & = & -19.28 \pm 0.35 ~ \rm mas~yr^{-1}, \nonumber
\end{eqnarray}
The error in this proper motion comes almost completely from the error
in measuring the positions of the QSO.  It is somewhat larger than the
error in a well-exposed star position.

Since QSOs behind other globular clusters will be useful too, it is of
interest to ask what positional accuracy can be expected for them.  We
believe that the answer is that they should have  the same accuracy as
stars.  The lower accuracy in the present case arises from an error in
the first-epoch position  that was 0.046   pixel, compared with  0.020
(just like a well-exposed star) in the second  epoch.  We believe that
this  error  comes from the  fact  that the QSO   is about 1 magnitude
fainter  than what  we would  consider well-exposed  and from the fact
that a diffraction spike was only 4 pixels away at  the first epoch (5
pixels at the second epoch).

(This spike  {\it corresponds} to the  one visible in  Fig.\ 5  to the
left of the    QSO, but on   account  of  a  difference  in  telescope
orientation  it was considerably closer   to the QSO  in the two-epoch
WFPC2 images on which our measurements were necessarily made.)

This proper motion was obtained only from the data sets F814W 1995 and
F814W  2000 deep.  The  other data-set pair  available for this object
(F336W 1995 and F814W 2000  shallow) was taken with different filters,
and this fact might introduce systematic biases.  Moreover, the latter
sets of images have significantly lower signal-to-noise ratio.

The error has been calculated as the sum, in quadrature, of the errors
in each epoch, and also an estimate of the error in the transformation,
based on the number of stars used in the local transformation and on the
rms of the displacement of these (Anderson \& King, in preparation).

Our proper motion is in agreement with that of Dinescu et al.\ (1999),
$\rm  (\mu_{\alpha}\cos\delta$,  $\rm \mu_\delta)_{\rm  J2000}$    $=$
$(-12.50\pm0.36,-19.93\pm0.49)$, in which they use  Hipparcos-measured
stars to  link relative motions to  an absolute system; our difference
is $1.4\sigma$ in the RA direction  and $1.1\sigma$ in the declination
direction.   As we  have said,  this agreement  with the solidly based
Hipparcos  system   strengthens our  identification of  our  reference
object as extragalactic.

In Galactic coordinates our proper motion becomes:
\begin{eqnarray}
(\mu_\ell \cos b)_{{\rm J2000}} & = & -23.30 \pm 0.35 ~ 
{\rm mas~yr^{-1}}, \nonumber \\
\mu_{b,{\rm J2000}}  & =  & -1.81 \pm 0.35 ~ {\rm mas~yr^{-1}}. \nonumber
\end{eqnarray}
In Fig.\ 6 we show the vector-point diagram of the proper
motions in Galactic coordinates, after placing the origin at the
extragalactic source.

\begin{figure}
\epsscale{1.0}
\plotone{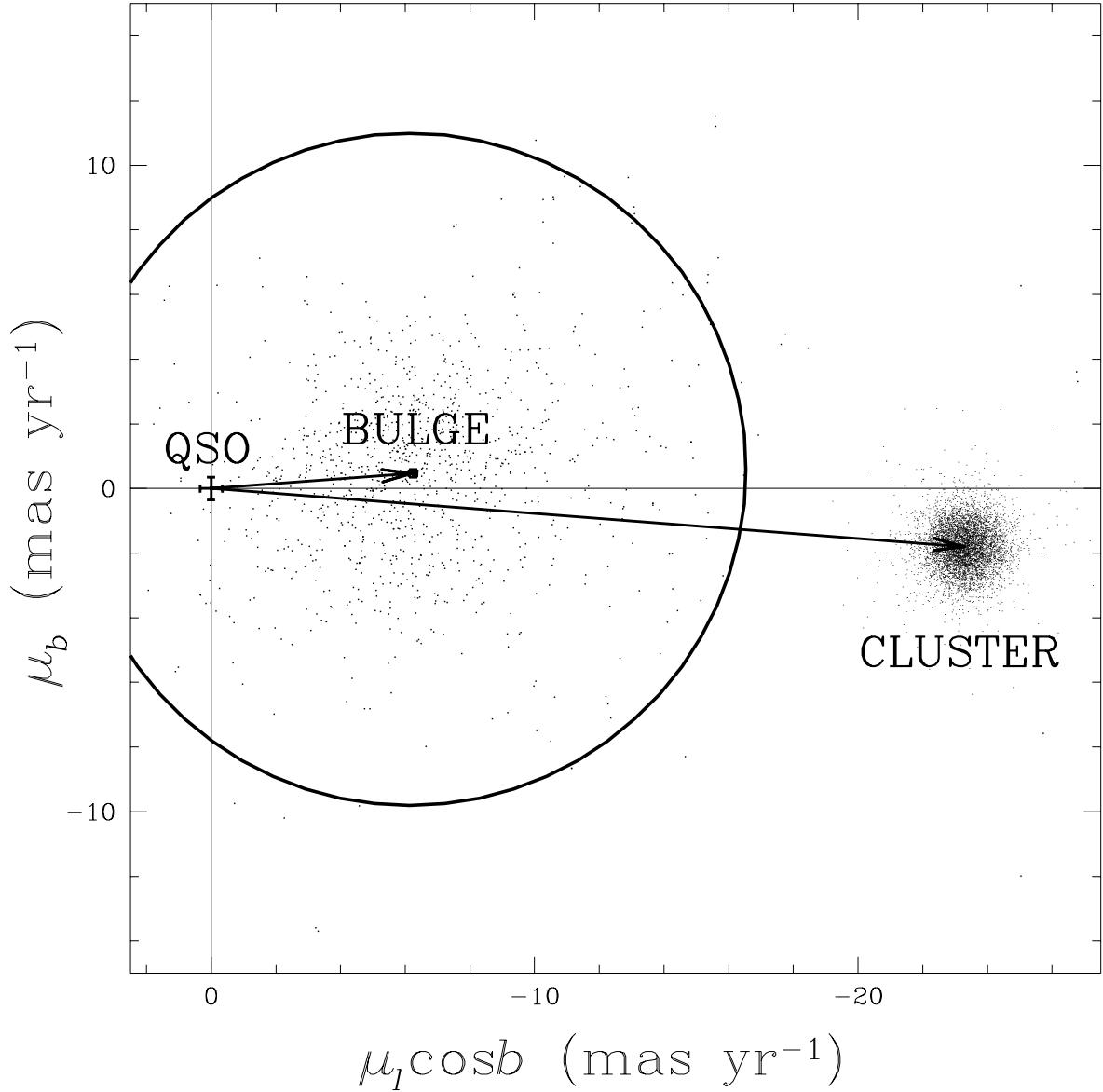}
\caption{ In this figure we show the proper motions of the bulge stars,
and M4 stars, in Galactic coordinates.  The extragalactic point source
has been chosen as the origin.  The meaning  of the circle is given in
the text. \label{bulge} }
\end{figure}

The mean  displacement  of the  cluster stars   with respect  to   the
extragalactic source is the combined effect of the  motions of the Sun
and the cluster around the Galaxy.  We use the computational matrix in
Johnson \& Soderblom (1987) to  derive the cluster space velocity with
respect  to the Sun,   $(U,V,W)_{{\rm  heliocentric}}$.   [Note   that
$(U,V,W)$ is a right-handed    coordinate system, so  components   are
positive in the direction  of the Galactic center, Galactic  rotation,
and the north Galactic Pole, respectively.]

Assuming   the following:  1) mean  coordinates  $(\alpha,\delta)_{\rm
J2000}$ = ($245^\circ.898,-26^\circ.525$), for the M4 fields; 2) J2000
Galactic coordinate system defined as before; 3) a distance of M4 from
the Sun of 1.8 kpc (a mean of  various values in the literature), with
an uncertainty  of 10\%;  4) a radial   velocity of 70.5 $\pm$  0.3 km
s$^{-1}$ (from  the Harris [1996] on-line  catalog, June 1999 update);
5) a right-ascension proper  motion $\mu_\alpha\cos\delta =  -13.21\pm
0.35$ mas yr$^{-1}$,  and a declination proper  motion of  $-19.28 \pm
0.35$ mas yr$^{-1}$ for M4, we get:
$ (U,V,W)_{{\rm heliocentric}}=(43\pm3,-207\pm20,-7\pm4$) km s$^{-1}$.

Then, assuming  a Galactocentric distance  for the Sun  of 8.0 kpc and
adopting      a     solar motion   of   $(U_\odot,V_\odot,W_\odot)   =
(10.0,5.2,7.2)$  km s$^{-1}$ (Binney \& Merrifield  1998, p.\ 628), we
get for M4
$$(U,V,W) =  (53\pm3,-202\pm20,0\pm4)  ~{\rm km~s^{-1}},$$  
which is the cluster space velocity relative  to the Local Standard of
Rest (LSR).

If we assume a LSR circular  motion of $V_0~=$  220 km s$^{-1}$ at our
distance of  8.0 kpc, the  cylindrical velocity components  of M4 in a
Galactic          rest              frame        are    $$(U,V,W)_{\rm
absolute}=(54\pm3,16\pm20,0\pm4) ~{\rm km }s^{-1},$$
which corresponds to a Galactic-rest-frame speed of 56 $\pm$ 20 km
s$^{-1}$.

We  note  how  a   small  variation  in  the assumed   parameters---in
particular in the dis\-tance---well within the uncertainties, can turn
the  orbit from prograde into   retrograde.  (If we   were to assume a
distance of 2.2 kpc for M4---as  in the on-line catalog of Harris---we
would get
$V \simeq -28\pm24$ km s$^{-1}$.)

Our measure of the absolute proper motion of M4 is not affected by its
annual parallax (which is $\sim0.5$ mas),  given that the observations
used in the present work were all taken at the same time of year.

\section{MEASUREMENT OF THE GALACTIC CONSTANT $V_0/R_0$}

M4 is a globular cluster projected near the edge of the Galactic bulge
($\ell\simeq-9^\circ$,  $b\simeq16^\circ$), at  a distance  of about 2
kpc from the Sun.  We expect a modest number  of foreground disk stars
in our fields (from here on we use ``foreground''  to mean in front of
the bulge rather than in front of the  cluster), but in the background
we look through the outer edge of the bulge, and the inner part of the
halo, at a height of  $\sim2$ kpc above  the plane.  Although at  such
heights the density of  the bulge and halo   are both rather low,  the
volume we are probing is  sizable, so that  we see  a large number  of
these stars.  Their absolute proper  motion is just the reflection  of
the Sun's angular  velocity with respect to  that point;  from that we
can derive the  value of the angular  velocity of the LSR with respect
to the Galactic center,  which  is the fundamental   Galactic-rotation
constant  $A-B=V_0/R_0$ (cf.\ Kerr    \& Lynden-Bell 1986,  Olling  \&
Merrifield 1998).  In this respect we do not need to involve ourselves
with  the complicated   question of whether  the   distant field stars
(after elimination of foreground stars) belong to the bulge, the halo,
or some transition population; the geometry along our line of sight is
the same in any case, with the average velocity  of these stars having
no component toward or away from the Galactic center.

To derive the value  of $V_0/R_0$ we need three  steps:\ (1)  find the
mean distance of the stars whose motion we  are observing, (2) correct
the observed proper motion for the velocity of the Sun with respect to
the LSR,  and (3) relate  the  corrected proper motion  to the angular
velocity of the LSR with respect to the Galactic center.

For the distance of the stars that we are observing, we adopt the
following working hypotheses:
\begin{enumerate}
\item The field stars are mainly bulge/halo members.  (We will show
below, on the basis of the observed distribution of the proper motions
and colors, that the foreground stars have a negligible influence.)
\item The stars whose motion we are studying are part of a spherical
spatial distribution around the Galactic center.  We use this hypothesis
just for simplicity in the estimation of the distance of its centroid; a
modest flattening (like that of the bulge) makes very little difference
in this.
\item Our observations go deep enough that we do not lose stars on the
far side of our sample, so that on the average  they are indeed at the
tangent point.
(In asserting this we neglect the small  effect of the $r^2$ flare-out
of our  cone of observation, which puts  the average distance a little
beyond  the tangent  point.    In the non-rotating   halo this  has no
effect, and in the  modestly rotating bulge  the stars in front of the
tangent   point and those behind   it,  closely equal  in number, have
oppositely directed transverse motions, which cancel each other in the
mean.)
\end{enumerate}

From these  it follows   that  we can   express  the distance  of  the
centroid, in  our line of sight, of  the  bulge/halo stars (which, for
brevity, we will  refer to  simply   as the  bulge) as a   geometrical
constant times the distance of the Sun from the Galactic center.
This distance is
\begin{equation}
R~=~R_0~\cos\ell\cos b.
\label{eq:mu1}
\end{equation}
If we take $R_0=8.0$ kpc, then $R=7.6$ kpc.

Next, the difference between the proper motion of  the bulge stars and
that of the extragalactic  point source is  the absolute proper motion
of that point in the bulge.  To measure  that motion, however, we must
remove the foreground  stars from our   sample of field stars.   To do
this we draw a putative halo main sequence by shifting that of M4 down
by 3.1  magnitudes, to allow for  the difference in distance.  Then we
draw a line 0.5 magnitude redder  in color, to  allow for the presence
of stars of higher metallicity and for a small amount of observational
error.  We exclude the 411 stars to the right of this line.

Fig.\ 6 shows the proper motions of M4 and of the surviving 1086 field
stars, in  Galactic  coordinates.   The  origin has  been set  at  the
extragalactic  point  source, labeled QSO.   The  error bars  show the
uncertainties  in the motions  with respect to  the mean motion of the
cluster stars.  We have  drawn a heavy circle  at a radius of 4  times
the  semi-interquartile distance of the field  stars from their median
position.  We considered stars inside it to be reliable members of the
population whose   motion we are studying,  so  we calculated the mean
motion from them.    This is the   mean absolute proper motion of  the
bulge, and is shown as an arrow in Fig.\ 6.  Its error---calculated as
$\sigma / \sqrt{{\rm Number~of~stars   ~used}}$---is indicated on  the
head of the arrow with a tiny error bar.  (The larger error bar at the
foot of the  arrow shows the measurement  error in the position of the
QSO.)  The  results  are:\  $\mu_\ell \cos  b =  -6.25  \pm  0.09$ mas
yr$^{-1}$, $\mu_b =  +0.47 \pm 0.08$ mas  yr$^{-1}$.  Inclusion of the
error in   the QSO position  increases the  sigmas  to $\pm  0.36$ mas
yr$^{-1}$, however.

We also carried out the  same procedure with the  entire set of  field
stars.  The resulting mean  motion differed by a  little less than 5\%
from what we got from the purified sample, showing that the foreground
stars  did indeed  need to be  removed  but  that our  results are not
sensitive to exactly how we did the rejection.

The apparent motion of the field stars is the  reflection of the Sun's
motion with respect to that point.  What we  need instead, however, is
the velocity  of the  LSR with  respect to our  region of observation;
this requires a  correction for the  solar motion with respect to  the
LSR.

Both for this step and  for the final  step  of converting the  proper
motion in this field into the apparent motion that we would observe at
the  Galactic center,  we  need to rotate the  components  of a vector
$(U,V,W)$,  in the local Galactic   system,  into an orientation  that
accords with the direction  of M4 from us,  so that $U_{\rm cl}$ is in
the radial-velocity direction,  $V_{\rm cl}$ in  the $\ell$ direction,
and  $W_{\rm cl}$   in the  $b$  direction.  That  rotation  is easily
accomplished in two steps.  First,  rotate around the north  celestial
pole so that the new $U_1$ axis points  toward longitude $\ell$ rather
than the Galactic center:
\begin{eqnarray}
U_1 &=& \phantom{-}U\cos\ell + V\sin\ell \nonumber \\
V_1 &=& -U\sin\ell + V\cos\ell \nonumber \\
W_1 &=& \phantom{-}W. \nonumber 
\end{eqnarray}
Then rotate  around the $V_1$ axis by  an angle  $b$, in the direction
from $U_1$ toward $W_1$:
\begin{eqnarray}
U_{\rm cl} &=& \phantom{-}U_1\cos b + W_1\sin b \nonumber \\
V_{\rm cl} &=& \phantom{-}V_1 \nonumber \\
W_{\rm cl} &=& -U_1\sin b + W_1\cos b. \nonumber 
\end{eqnarray}
The combination of the two is
\begin{eqnarray}
U_{\rm cl} &=& \phantom{-}(U\cos\ell + V\sin\ell)\cos b + W\sin b
\nonumber \\ 
V_{\rm cl} &=& -U\sin\ell + V\cos\ell \\
W_{\rm cl} &=& -(U\cos\ell + V\sin\ell)\sin b + W\cos b, \nonumber 
\end{eqnarray}

The effect of the Sun's  motion with respect  to the LSR, for which we
assume $(U,V,W)   = (10.0,5.2,7.2)$ km  s$^{-1}$(Binney  \& Merrifield
1998, p.\  628), then  becomes  $V_{\rm cl}=6.7$ km  s$^{-1}$, $W_{\rm
cl}=4.4$ km s$^{-1}$.  The  corresponding corrections to  our observed
proper  motion  of the  bulge are these  linear  velocities divided by
$4.74R$  (where  4.74  is   the equivalent   in  km  s$^{-1}$  of  one
astronomical unit  per tropical year, and  expressing $R$ in  kpc will
give a result in  mas  yr$^{-1}$).  Thus $\Delta\mu_\ell\cos  b=0.186$
mas   yr$^{-1}$, $\Delta\mu_b=0.123$ mas   yr$^{-1}$.  Since it is the
reflection of this that shows up at  the bulge, we must subtract these
quantities from the observed proper motion of the bulge.

The resulting angular proper motion of our bulge field with respect to
the LSR  is  thus $\mu_\ell \cos b   = -6.06 \pm 0.36$  mas yr$^{-1}$,
$\mu_b =  +0.35 \pm  0.36$   mas yr$^{-1}$.  (Note  that because   the
correction for Solar motion is a small one, [1]  our choice of a value
for $R_0$  was not critical  and [2] we  need not increase  our quoted
errors on account of any uncertainty in the Solar motion.)

For the final step of moving from  the apparent proper motion of bulge
stars at the position of M4 to the  angular velocity of the LSR around
the Galactic center, we note that if we assume circular motion for the
LSR, then its velocity  $V_0$ is perpendicular  to a line from the Sun
to the Galactic center, so  that $(U,V,W)=(0,V_0,0)$.  Eqs.\ (2)  then
give
\begin{eqnarray}
V_{\rm cl} &=& \phantom{-}V_0\cos\ell \nonumber \\
W_{\rm cl} &=& -V_0\sin\ell\sin b. \nonumber 
\end{eqnarray}

On  the  other hand,  the  relation between  transverse velocities and
proper motions gives, with substitution from Eq.\ (1),
\begin{eqnarray}
V_{\rm cl} &=& -4.74R(\mu_\ell\cos b) \nonumber \\
           &=& -4.74R_0\cos l\cos b(\mu_\ell\cos b) \nonumber \\
W_{\rm cl} &=& -4.74R\mu_b \nonumber \\
           &=& -4.74R_0\cos l\cos b\mu_b. \nonumber 
\end{eqnarray}
Combining these two pairs of equations then gives
\begin{eqnarray}
{1\over\cos b}{V_0\over R_0} &=& -4.74(\mu_\ell\cos b) \nonumber \\
\tan l\tan b{V_0\over R_0}   &=& \phantom{-}4.74\mu_b. \nonumber 
\end{eqnarray}
These   are  two  equations  for $V_0/R_0$,    which  give independent
estimates of its value:
\begin{eqnarray}
{V_0\over R_0} &=& -4.74\cos b(\mu_\ell\cos b) \nonumber \\ 
{V_0\over R_0} &=& \phantom{-}4.74{\mu_b\over\tan l\tan b}. \nonumber 
\end{eqnarray}
If   we were to take  a  weighted mean  (or  what  is equivalent, find
$V_0/R_0$ by  least   squares  from the  two equations),   the  second
equation would make only a very small contribution.  Since the referee
has  pointed  out to   us, besides,  that  $\mu_b$   would need  to be
corrected for a systematic contribution from bulge rotation seen at an
oblique angle---a very  uncertain quantity---we  choose to solve   for
$V_0/R_0$   from the $\mu_\ell\cos b$ equation   alone.  The result is
$V_0/R_0=27.6
\pm 1.7$ km s$^{-1}$ kpc$^{-1}$.

This  is an independent measurement of   this quantity, of which there
are various  other  values in  the  literature.    It is   related  in
complicated ways to  the other constants  of Galactic  rotation, and a
full discussion of that problem would be far  beyond the scope of this
paper.  Suffice it to say that our  value is close to the $26.4\pm1.9$
km s$^{-1}$ kpc$^{-1}$  that is quoted  in the comprehensive review by
Kerr \& Lynden-Bell (1986).

Although we have argued that our results  do not depend on whether our
field  stars are members of the  Galactic bulge or halo population, we
conclude this section by presenting some evidence that relates to this
question, in  the form of  the sizes of the  velocity dispersions.  We
took each of these to  be half of the  interval, centered on the mean,
that includes 68.3\% of the sample.  The results are
\begin{eqnarray}
\sigma_{\mu_\ell\cos b} &=& 2.99 \pm 0.09 ~{\rm mas~yr^{-1}}, \nonumber \\
\sigma_{\mu_{b}}        &=& 2.63 \pm 0.08 ~{\rm mas~yr^{-1}},  \nonumber
\end{eqnarray}
corresponding  to 108  and  95 km  s$^{-1}$,  respectively,  under the
assumption   of  $R =  7.6$  kpc  (corresponding to   $R_0 = 8.0$ kpc,
according to  Eq.\ \ref{eq:mu1}).  For what  it is worth, these values
correspond  to those found  by Spaenhauer, Jones,  \& Whitford (1992),
and by Kuijken \& Rich (2002), for bulge fields closer to the Galactic
center.  The higher dispersion  along  Galactic parallels than   along
Galactic meridians is suggestive of Galactic rotation, which should be
seen in the bulge but not in the halo.

\section{CONCLUSIONS}
\label{conc} 

In  this work we  have presented  an  astrometric study of the closest
Galactic globular cluster M4, based on multi-epoch \hst\ observations.
We have been  able to separate  almost perfectly the member stars from
the field  objects,  and  to identify  an   extragalactic point-source
candidate (QSO).
The QSO allowed us  to measure the  absolute proper motion of  M4, for
which  we  find $(\mu_\alpha\cos\delta,~\mu_{\delta})_{\rm J2000}$ $=$
$( -13.21 \pm 0.35, -19.28 \pm 0.35) ~ \rm mas ~ yr^{-1}$.
Moreover, we have been able  to measure the  apparent proper motion of
the bulge and  from it to  derive an estimate of the  value of $A-B  =
V_0/R_0 = 27.6 \pm 1.7$ km s$^{-1}$ kpc$^{-1}$.

In a future  paper  we will  use these  proper  motions to study   the
cluster members in more detail, studying the color--magnitude diagram,
two-color  diagram,  luminosity function,   mass  function, and   mass
segregation.  We are also working on the full data base, including the
new ACS/WFC images, to derive internal proper motions.  It should soon
be possible to derive a precise geometrical distance (with an accuracy
of a few  per cent)  based  on the   comparison  of the  proper-motion
dispersion with  that of the  radial velocities for  a large sample of
stars, using newly available multi-fiber high-resolution spectroscopic
facilities  like FLAMES$+$GIRAFFE at $VLT$,  a project we have already
submitted to ESO.

%
%
\acknowledgments We are grateful to Andrea Grazian for providing us
simulated two-color  diagrams of QSOs  in \hst\ filters, and to Harvey
Richer  for  pointing  out   that we   needed   to consider   possible
contamination by foreground stars.   This  project has been  partially
supported by the Ministero dell'Istruzione e della Ricerca Scientifica
and   by the  Agenzia  Spaziale  Italiana.  I.\  R.\  K.\ and  J.\ A.\
acknowledge support from STScI Grant GO-8153.

%
%
\newpage     
%

%
%

\end{document}